\documentclass[floatfix,aps,prb,twocolumn,showpacs]{revtex4}
\usepackage{graphicx}
\usepackage[]{epsfig}
\usepackage{times,amsmath,amssymb}
\begin{document}
\title{Detection of spin polarization with a side coupled quantum dot}

\author{Tomohiro Otsuka}
\email{t-otsuka@issp.u-tokyo.ac.jp}
\author{Eisuke Abe}%
\author{Yasuhiro Iye}%
\author{Shingo Katsumoto}%

\affiliation{Institute for Solid State Physics, University of Tokyo, 5-1-5 Kashiwanoha, Kashiwa, Chiba 277-8581, Japan}

\date{\today}
\begin{abstract}
We propose realistic methods to detect local spin polarization, which utilize
a quantum dot side coupled to the target system.
By choosing appropriate states in the dot, we can put spin selectivity to the dot and detect spins in the target with small disturbance.  
We also present an experiment
which realizes one of the proposed spin detection schemes in magnetic fields.
\end{abstract}

\pacs{73.63.Kv, 73.63.Nm, 72.25.Dc, 85.35.-p}
\maketitle
\section{Introduction}
Electrical generation of electron spin polarization in non-magnetic semiconductors is a key technology in the development of spintronics~\cite{2001WolfSci, 2004ZuticRMP}.
The use of spin-orbit interaction~\cite{1960RashbaFTT, 1984BychovJPC, 1955DresselhausPR} enables spin polarization without ferromagnets or external magnetic fields, and a variety of forms of spin filters based on the interaction have been proposed~\cite{2001KiselevAPL, 2004PareekPRL, 2005OhePRB, 2005EtoJPSJ}.
Although most of the proposed devices are technologically feasible, the main obstacle to the experimental verification lies in the detection part.
A small ferromagnet attached to the outlet of a spin filter may be used as a polarization detector owing to its spin-dependent transparent coefficients.
However, at the semiconductor-metal interface, the polarization generally suffers from conductance mismatch~\cite{2000SchdmitPRB} and disorder-induced scattering, resulting in low detection efficiency and strong disturbance to the spin source.
It is also obvious that the use of ferromagnets is incompatible with the use of spin-orbit interaction. 
Therefore, polarization detectors that are made of the host material alone and ``quiet" to the spin source are highly desired.

In this article, we describe realistic methods to detect spin polarization by using a quantum dot tunnel-coupled to a target (such as a spin filter).
Since the dot and the target are connected at only a single point, our methods require no net current to flow through the dot, thus realizing a detector with extremely small disturbance to the target.
These are suitable for the detection of spin polarization not only for delicate spin filters but also for spin Hall effect~\cite{2005MurakamiASSP, 2006SchliemannIJMPB}.
Moreover, our methods are quite general and applicable to any system or material in which a few-electron quantum dot can be prepared.
As a proof of this principle, we present experiments on the polarization detection in magnetic fields utilizing a quantum wire as a controllable spin polarization source.

\section{Detection principle}
\begin{figure}[b]
\begin{center}\leavevmode
\includegraphics{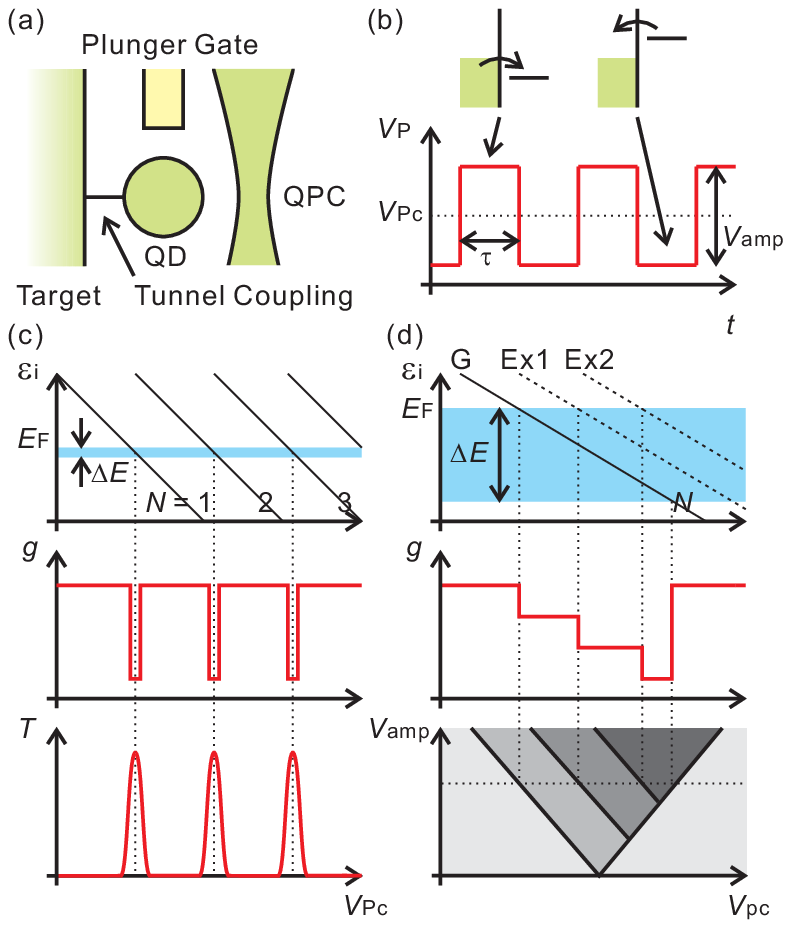}
\caption{(Color online) (a) Schematic of the proposed spin polarization detector.
(b) Waveform of $V_{\rm P}$ to perform energy spectroscopy in the side coupled QD.
Upper figures show the movement of an energy level in QD in phase with $V_{\rm P}$.
(c) Shift of $\epsilon_{i}$ (upper), schematic of $g$ (middle), and Coulomb oscillations in the transmission probability $T$ if we prepare a single electron transistor structure (lower).
The lines in the upper figure show the energy levels of $N=1,~2$ and 3 ground states from left to right.
The energy window is indicated by the gray (blue) zone.
(d) Shift of $\epsilon_{i}$ (upper), schematic of $g$ with large $V_{\rm amp}$ (middle), and a gray scale plot of $g$ as a function of $V_{\rm Pc}$ and $V_{\rm amp}$ (lower).
The lines in the upper figure correspond to the energy of ground, first and second excited levels of $N$-electron  state from left to right.
}
\label{Fig1}
\end{center}
\end{figure}

Figure 1(a) shows the setup of the detector, in which a quantum dot (QD) is tunnel-coupled to a target device via a single contact.
The electrostatic potential of QD is detected through the conductance of
a quantum point contact (QPC) placed on the other side of QD~\cite{1993FieldPRL, 1998BuksNat, 2003ElzermanPRB}.
We first explain how the energy spectrum of QD (including  excited states)
is obtained in this device~\cite{2004ElzermanAPL, 2008OtsukaAPL}.
For the spectroscopy, the plunger gate voltage $V_{\rm P}$ on QD is
driven to oscillate as a rectangular wave with the amplitude $V_{\rm amp}$
and the center voltage $V_{\rm Pc}$, as illustrated in Fig.~1(b).
$V_{\rm P}$ shifts the energy levels in QD up and down, and induces shuttling of an electron between QD and the target.
When the shuttling is forbidden, the electrostatic potential of QD simply follows the oscillation of $V_{\rm p}$ and thus the conductance through QPC results in a square wave in phase with $V_{\rm P}$.
When the shuttling has a finite probability, an electron injected into QD screens some portions of the variation of the electrostatic potential, and diminishes the oscillation of the conductance.
Then the shuttling is detected by the decrease in a lock-in signal $g$ of the conductance through QPC in phase with the applied square wave.
From the diminishment of the lock-in signal $\Delta g$, the information on the energy levels in QD can be extracted.

An analytical approach tells that
\begin{equation}
\Delta g\propto1-\frac{\pi^2}{\Gamma^2\tau^2+\pi^2},
\label{eq_depth}
\end{equation}
where $\Gamma$ is the tunneling rate and $\tau$ is the half period of the square wave~\cite{2004ElzermanAPL}.
$\Gamma$ is formally written as
\begin{equation}
\Gamma=\sum_{E_{\rm F}-\Delta E<\epsilon_{i}<E_{\rm F}}\gamma_i,
\end{equation}
where $i$ is the level index, $\gamma_i$ and $\epsilon_{i}$ are the coupling
constant and the energy level in QD at injection process (the period during $V_P=V_{Pc}+1/2V_{\rm amp}$), respectively.
$E_{\rm F}$ is Fermi energy of the target device and $\Delta E=e(C_{\rm g}/C)V_{\rm amp}$ is the width of the energy window in which the shuttling occurs.
Levels satisfying $E_{\rm F}-\Delta E<\epsilon_{i}<E_{\rm F}$ cross $E_{\rm F}$ by the oscillation with $V_{\rm amp}$.
In the simplest case, $\gamma_i$ is a common constant $\gamma_0$ and $\Gamma$ becomes $N_{\rm in}\gamma_0$, where $N_{\rm in}$ is the number of levels in the energy window.
When $V_{\rm Pc}$ is swept, the energy levels in QD cross the energy window and $\Gamma $ changes with the change of $N_{\rm in}$.
When $\Delta E$ is narrower than the energy level spacing, $N_{\rm in}$ takes the value of 0 or 1 with the change of $V_{\rm Pc}$.
Then the energy levels in QD appear as a series of dips in
$g$, the positions of which give the energy spectroscopy of the ground states in QD similar to the Coulomb oscillation (Fig.~\ref{Fig1}(c)).
If we increase $V_{\rm amp}$ and widen $\Delta E$, $N_{\rm in}$ can be larger than 1.
In this case, $g$ should show stepwise decreases at which excited states come into the window as shown in Fig.~\ref{Fig1}(d).
Hence the excitation energy spectra with fixed number of electrons in QD $N$ can be obtained from the line shape of $g$ versus $V_{\rm Pc}$
with a large $V_{\rm amp}$.
Here we assume that the dips are well separated, {\it i.e.}, the charging
energy is much larger than energy level spacing between the excited states and multiple occupation of QD is forbidden.

\begin{figure}[]
\begin{center}\leavevmode
\includegraphics{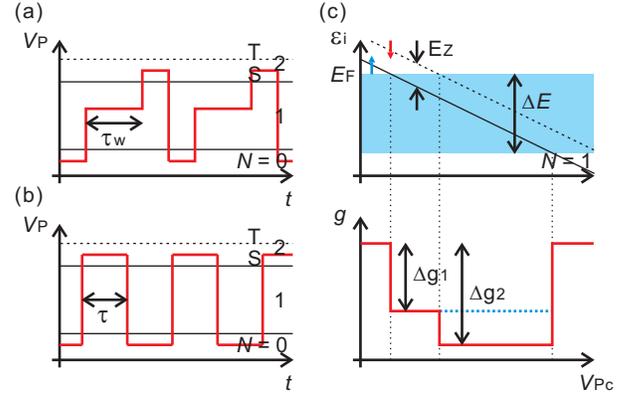}
\caption{(Color online) (a) Sequence of $V_{\rm P}$ to detect spin polarization. The horizontal lines show the energy levels of triplet and singlet of $N=2$ and $N=1$ states from top to bottom. 
(b) Another sequence of $V_{\rm P}$ to detect spin polarization. 
(c) Scheme of spin polarization detection in high magnetic fields. Shift of $\epsilon_{i}$ (upper), and schematic of $g$ (lower). 
The lines in the upper figure correspond to the energy levels of up- and down-spins of $N=1$ state from left to right.
The dotted trace in the lower figure shows the result for $P=1$.
}
\label{Fig2}
\end{center}
\end{figure}

Now we move onto the detection of spin polarization.
Let $D_{\uparrow}$ and $D_\downarrow$ be the density of states at $E_{\rm F}$ for up- and down-spins, respectively.
The spin polarization $P$ is defined as $(D_\uparrow-D_\downarrow)/(D_\uparrow+D_\downarrow)$.
To detect the polarization, we use two-electron state in QD.
Due to the Pauli principle, the ground state is spin singlet and the excited state is spin triplet.
%Detection with 2-step pulse
A way for utilizing this property for polarization detection is to apply the sequence of $V_{\rm P}$ illustrated in Fig.~\ref{Fig2}(a).
First, $V_{\rm P}$ is set in $N=0$ region and QD is emptied.
Secondly, the first electron is injected and wait time $\tau _{\rm w}$.
Finally, $V_{\rm P}$ is set between the singlet and triplet levels of $N=2$ state.
When we make $\tau_{\rm w}$ much longer than the spin relaxation time $T_1$, the spin state of the first electron is not conserved and both spin states are realized with the probability of 1/2.
Then to form spin singlet, the tunneling rate of the second electron becomes
\[
\Gamma=\frac{1}{2}AD_\uparrow+\frac{1}{2}AD_\downarrow=\frac{A}{2}(D_\uparrow+D_\downarrow)\equiv\Gamma_{\rm nc},
\]
where $A$ is a constant.
The first term represents the case in which the second electron has up-spin and the second term  shows the opposite case.
On the other hand, when $\tau_{\rm w}$ is much shorter than $T_1$, the spin state is conserved and the rate is
\[
\Gamma=
\frac{D_\downarrow}{D_\uparrow+D_\downarrow}AD_\uparrow +
\frac{D_\uparrow}{D_\uparrow+D_\downarrow}AD_\downarrow
=A\frac{2D_\uparrow D_\downarrow}{D_\uparrow+D_\downarrow}\equiv\Gamma_{\rm c},
\]
which is equal to or less than $\Gamma_{\rm nc}$.
Now $P$ can be obtained from the following relation,
\begin{equation}
\frac{\varDelta\Gamma}{\Gamma_{\rm nc}}
\equiv\frac{\Gamma_{\rm nc}-\Gamma_{\rm c}}{\Gamma_{\rm nc}}
=\frac{(D_\uparrow-D_\downarrow)^2}{(D_\uparrow+D_\downarrow)^2}
=P^2.
\end{equation}
This equation allows us to evaluate $P$ from the measurement of $\Gamma $.
%Detection with 2-step wave

%Detection with square wave
In another method to detect the change of spin polarization, we utilize the square wave illustrated in Fig.~\ref{Fig2}(b).
$V_{{\rm amp}}$ and $V_{{\rm Pc}}$ are adjusted so that $N=1$ and the singlet level of $N=2$ states are contained in the energy window.
At injection phase, two electrons can be injected into QD.
The tunneling rate of the second electron is again given as
\begin{equation}
\Gamma
=A\frac{2D_\uparrow D_\downarrow}{D_\uparrow+D_\downarrow}.
\end{equation}
Here we assume $\tau $ is much shorter than $T_1$.
With the change of $P$, $\Gamma $ is modified and this results in $\Delta g$.
Especially if $P=1$ ($D_{\downarrow} = 0$), it vanishes and the injection of the second electron is forbidden.
With this technique, we can confirm the change of $P$ in spin filters which have controllability with external parameters~\cite{2005EtoJPSJ, 2008AharonyPRB}.
%Detection with square wave
The above two schemes work even in zero-magnetic field.% and the net current through QD during the measurement is zero.

%Detection in high magnetic field
Under the magnetic field, the detection becomes simpler.
We again apply square wave on $V_{\rm P}$.
By applying magnetic field, the lowest level of $N=1$ state can accept only an up-spin electron due to the Zeeman energy $E_{\rm Z}$ as illustrated in Fig.~\ref{Fig2}(c).
When only this state is in the energy window, the rate becomes $\Gamma=AD_\uparrow\equiv\Gamma_1$.
With increasing $V_{\rm Pc}$, the upper Zeeman state comes into the energy window and the rate changes to $\Gamma=A(D_\uparrow+D_\downarrow)\equiv\Gamma_2$.
The ratio is thus
\begin{equation}
\frac{\Gamma_1}{\Gamma_2}=\frac{D_\uparrow}{D_\uparrow+D_\downarrow}=\frac{1+P}{2},
\label{eq_ratio_undermag}
\end{equation}
which takes the value from 1/2 to 1.
The former and the latter correspond to $P=0$ and $P=1$, respectively.
$\Gamma_1$ and $\Gamma_2$ are obtained from the two depths  $\Delta g_1$ and $\Delta g_2$ illustrated in Fig.~\ref{Fig2}(c).
In this method, $T_1$ has no effect on the injection of an electron. 
In the following, we present experimental demonstration of this detection method.
%Detection in high magnetic field

\section{Experimental demonstration}
\begin{figure}[]
\begin{center}\leavevmode
\includegraphics{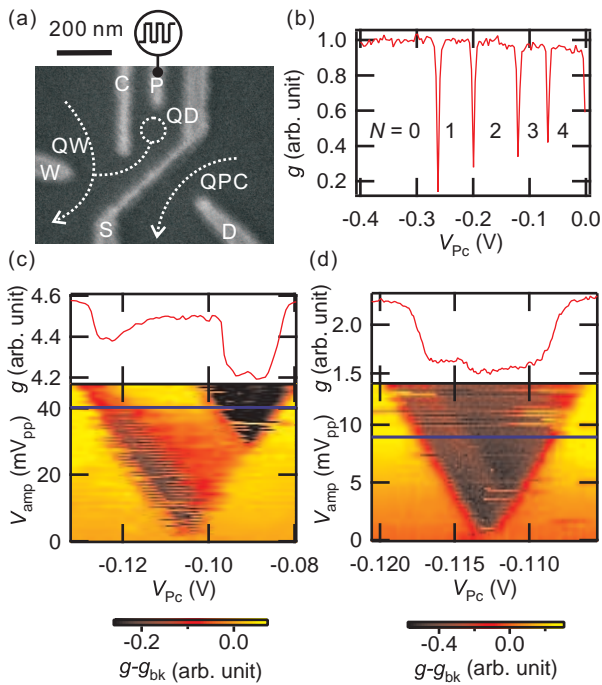}
\caption{(Color online)
(a) Scanning electron micrograph of the device. 
White regions are Au/Ti gates deposited on the surface of a GaAs/AlGaAs wafer.
(b) Measured $g$ as a function of $V_{\rm Pc}$ with $V_{\rm amp}=4.5$~mV and $\tau =1053~\mu$s. 
(c) $g$ as a function of $V_{\rm Pc}$ and $V_{\rm amp}$  with  $\tau =680~\mu$s (lower).
In the graph, the background $g_{\rm bk}$ increasing with $V_{\rm amp}$ is subtracted for clarity.
The upper graph shows the cross section along the solid line.
(d) $g$ as a function of $V_{\rm Pc}$ and $V_{\rm amp}$ (lower) and the cross section (upper).
Magnetic field 14~T is applied in parallel to the 2DEG. 
}
\label{Fig3}
\end{center}
\end{figure}

Figure~\ref{Fig3}(a) shows a scanning electron micrograph of the device
to realize the setup shown in Fig.~\ref{Fig1}(a).
Au/Ti Schottky gates are deposited on a GaAs/AlGaAs heterostructure wafer containing two-dimensional electron gas (2DEG, depth: 60~nm, carrier density: $2.1 \times 10^{15}$~m$^{-2}$, mobility: 32~m$^{2}$/Vs).
By applying negative voltages on gates S, P, C and W, QD side coupled to a quantum wire (QW) is formed.
QW is the target of the spin polarization detection in this setup.
QPC is formed by gates S and D, and acts as the detector of the electrostatic potential of QD.
$g$ is detected with lock-in measurement of the conductance through QPC in phase with the applied square wave on gate P.
$\tau$ was set as 600~$\mu $s in almost all measurements.
The device was cooled down to 30~mK by a dilution refrigerator, and magnetic field up to 14~T in parallel to the 2DEG plane was applied by a superconducting solenoid.

In Fig.~\ref{Fig3}(b), (c), (d), we review energy spectrum measurement reported in Ref.~\onlinecite{2008OtsukaAPL}.
When $V_{\rm amp}$ is small, $g$ versus $V_{\rm Pc}$ shows a series of
dips like Fig.~\ref{Fig1}(c) (Fig.~\ref{Fig3}(b)), giving the ground state spectrum.
From this signal we can determine $N$.
With increasing $V_{\rm amp}$, the dip widens (Fig.~\ref{Fig3}(c)).
A stepwise drop of $g$ due to the first excited orbital also appears.
Application of magnetic field parallel to the 2DEG causes the Zeeman splitting of spin states, which is detected by the drop of $g$ (Fig.~\ref{Fig3}(d)).
Note that in this measurement the one dimensional bands in QW are also split
by the Zeeman effect but $E_{\rm F}$ is placed far from the band edges and
the difference between $D_\uparrow$ and $D_\downarrow$ can be ignored.

Now let us turn to the spin polarization detection.
Ideally the conductance of QW under zero magnetic field is quantized in units of $2e^2/h$, where the factor 2 originates from Kramers degeneracy.
This produces staircase-like variation of the conductance through QW $G_{\rm W}$ versus the width of QW controlled by the wire gate voltage $V_{\rm W}$~\cite{1988vanWeesPRL}.
Application of the parallel magnetic field lifts the Kramers degeneracy and
conductance plateaus at odd multiples of $e^2/h$ appear~\cite{1988vanWeesPRB, 1988WharamJPC}.
On such intermediate conductance plateaus, $E_{\rm F}$ lies between the two Zeeman-separated edges of one-dimensional bands, therefore, $D_\uparrow$ and $D_\downarrow$ are different.
For example, at the lowest conductance plateau, only an up-spin band is occupied ($P=1$).
The difference becomes very small on the conductance plateaus with even multiples
of $e^2/h$.
Hence $P$ oscillates against $V_{\rm W}$.

\begin{figure}[]
\begin{center}\leavevmode
\includegraphics{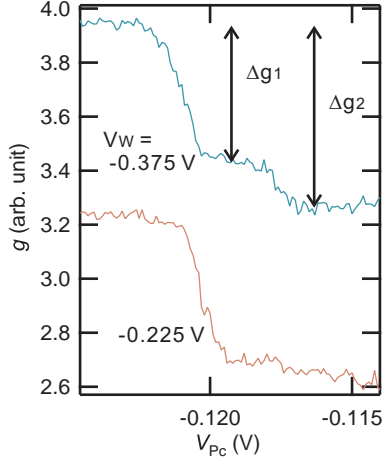}
\caption{(Color online)
$g$ as a function of $V_{\rm Pc}$. The upper and lower traces are the results with $V_{\rm W}=-0.375$ and $-0.225$~V, respectively. 
The upper trace has offset 0.7 for clarity.
}
\label{Fig4}
\end{center}
\end{figure}

To detect $P$ in QW, we apply the third method described in the previous section.
Figure~\ref{Fig4} shows measured $g$ as a function of $V_{\rm Pc}$ with changing $V_{\rm W}$ at 14~T.
The upper trace is shifted by 0.7 for clarity.
A two-step dip structure like Fig.~\ref{Fig2}(c) is observed in the case of $V_{\rm W}=-0.375$~V.
But with $V_{\rm W}=-0.225$~V, the trace is single-step and indicates $P=1$.
In this measurement, the change of $\Gamma _1$ with the shift of $V_{\rm W}$ is compensated by the readjustment of the coupling gate voltage $V_{\rm C}$.

\begin{figure}[]
\begin{center}\leavevmode
\includegraphics{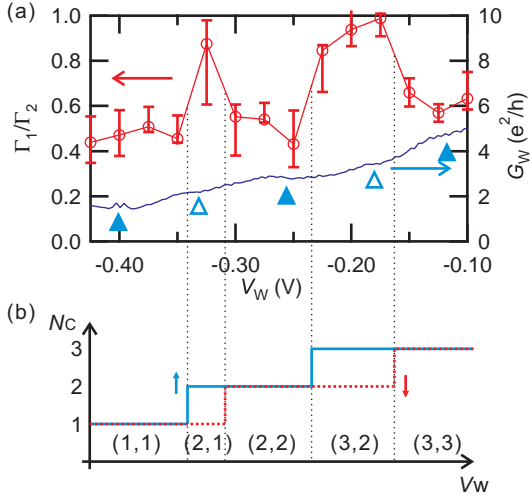}
\caption{(Color online)
(a) $\Gamma _1/\Gamma _2$ (open circles) and $G_{\rm W}$ (lower solid trace) as a function of $V_{\rm W}$.
Open and solid triangles indicate the positions of odd and even plateaus, respectively.
Error bars of $\Gamma _1/\Gamma _2$ arise from the uncertainties in evaluating $\Delta g$.
(b) Estimated $N_{\rm C}$ for up- (solid trace) and down-spins (dotted trace) as a function of $V_{\rm W}$.
$(N_{{\rm C}\uparrow }, N_{{\rm C}\downarrow})$ are indicated at the lower of the figure.
}
\label{Fig5}
\end{center}
\end{figure}

In Fig.~\ref{Fig5}(a), we plot $G_{\rm W}$ as a function of $V_{\rm W}$.
Short conductance plateaus (marked with open triangles) and long plateaus (solid triangles) are observed alternately.
The short and long plateaus correspond to odd ($N_{{\rm C}\uparrow }+ N_{{\rm C}\downarrow}=2n+1$) and even ($N_{{\rm C}\uparrow }+ N_{{\rm C}\downarrow}=2n$) plateaus because the Zeeman splitting is smaller than the spacing between one-dimensional bands.
Here, $N_{{\rm C}\uparrow }$ and $N_{{\rm C}\downarrow}$ are the number of channels with up- and down-spins, respectively, and $n$ is a non-negative integer.
The estimated $N_{{\rm C}}$ is shown in Fig.~\ref{Fig5}(b).
In this device, the value of $G_{\rm W}$ at plateaus are not the multiples of $e^2/h$ due primarily to the complexity of the gate configuration.

In Fig.~\ref{Fig5}(a), the values of $\Gamma_{\rm 1}/\Gamma_{\rm 2}$ obtained from the measured $\Delta g_1/\Delta g_2$ are shown with open circles.
Near the even plateaus of $G_{\rm W}$, $\Gamma_{\rm 1}/\Gamma_{\rm 2}$ is around 0.5 indicating $P$ is almost 0.
On the other hand, around the odd plateaus, $\Gamma_{\rm 1}/\Gamma_{\rm 2}$ goes up to 0.9 or 1.0 showing $P\approx 1$. 
This behavior is qualitatively in accordance with what is expected.
These results certificate that the system is working as a spin polarization detector.
Nevertheless, two puzzling points remain.
(1) The region with large $\Gamma_{\rm 1}/\Gamma_{\rm 2}$ at $(N_{{\rm C}\uparrow }, N_{{\rm C}\downarrow})=(3,2)$ is significantly wider than the width of the conductance plateau.
(2) $\Gamma_{\rm 1}/\Gamma_{\rm 2}$ goes up to 1 even at higher plateau.
If all of the channels have the same contribution on $P$, $P$ is represented by $(N_{{\rm C}\uparrow }-N_{{\rm C}\downarrow})/(N_{{\rm C}\uparrow }+N_{{\rm C}\downarrow})\leq 1/(N_{{\rm C}\uparrow }+N_{{\rm C}\downarrow})$.
Hence $P$ should decrease with increasing $N_{\rm C}$. 

As for the puzzle (1), the reason will be the spatial difference between the position at which $G_{\rm W}$ is determined and the one at which QD couples to QW.
Because QW in the present experiment is comparatively short, the electron wave function should be inevitably inhomogeneous along QW.
$G_{\rm W}$ is determined by the narrowest part of QW while QD detects $P$ at the connection point to QW.
The two points are not necessarily same.
Lowering of the band edge of higher channel down to $E_{\rm F}$ may occur at the QD-QW contact before the high channel opens at the narrowest point.
This makes the difference between $\Gamma_{\rm 1}/\Gamma_{\rm 2}$ and $G_{\rm W}$ around the condition on which a new channel opens.

\begin{figure}[]
\begin{center}\leavevmode
\includegraphics{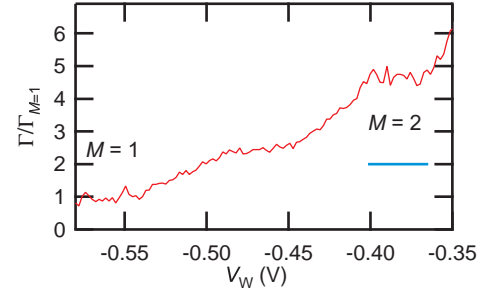}
\caption{(Color online)
Measured $\Gamma /\Gamma _{M=1}$ as a function of $V_{\rm W}$ at zero magnetic field.
The horizontal line around $V_{\rm W}=-0.38$~V shows the estimated $\Gamma /\Gamma _{M=1}$ assuming $M=1$ and 2 channels have same coupling strength.
}
\label{Fig6}
\end{center}
\end{figure}

As for puzzle (2), the larger coupling strength between QD and QW for the higher channel will be the reason.
To confirm this point, we check the change of $\Gamma $ when a new channel in QW opens.
Figure~\ref{Fig6} shows the result.
The measurement is performed in a device with the same design at zero magnetic field.
$\Gamma $ is normalized by the tunneling rate of the lowest channel $\Gamma _{M=1}$, where $M$ is an index for channels.
Steep increases of $\Gamma $ are observed around $V_{\rm W}=-0.51$ and $-0.42$~V which correspond to the beginning points of the ejection and injection with $M=2$ channel, respectively.
If $M=1$ and 2 channels have the same coupling strength, $\Gamma /\Gamma _{M=1}$ becomes 2 when both channels open around $V_{\rm W}=-0.38$~V as indicated with a horizontal line in Fig.~\ref{Fig6}.
But the observed result is apparently larger than this line and shows that the higher channel has larger coupling.
This originates from the shape of the wave function  across QW.
In the higher channel, the wave function of QW is extended to outer side and has larger overlap with the wave function of QD.

In the last part, we emphasize the small detection current needed in this detection scheme.
The net current through QD during the measurement is apparently zero.
The ``exchange current" between the target and QD, defined as $=e/2\tau $, is also extremely small.
In the detection in high magnetic field, we use $\tau\sim 1$~ms.
Then the exchange current is about 100~aA.
For the detection at zero magnetic field, because $T_1$ is the order of ms~\cite{2003HansonPRL,2004ElzermanNat}, $\tau $ should be smaller by an order or more.
The exchange current in this case is 1$\sim$10~fA, but still two orders of magnitude smaller than the current for ordinary conductance measurements.
From these, we expect the proposed detector is promising for sensing spin polarization created in delicate mechanisms such as spin filtering by spin-orbit interaction, spin Hall effect, etc.

In summary, we proposed realistic methods to detect local spin polarization by using a quantum dot side coupled to a target device.
With a quantum wire in magnetic fields as a controllable spin polarized source, we demonstrated the operation of the polarization detector experimentally.
In the case of a quantum wire, we found that the dot selectively senses polarization of electrons in a channel which has largest coupling to the dot.

We thank Y. Hashimoto for technical supports.
This work is supported by Grant-in-Aid for Scientific Research and Special Coordination Funds for Promoting Science and Technology.

\end{document}